# Nonlinear Damping of the 'Linear' Pendulum

Randall D. Peters
Department of Physics
Mercer University
Macon, Georgia

**ABSTRACT**
This study shows that typical pendulum dynamics is far from the simple equation of motion presented in textbooks. A reasonably complete damping model must use nonlinear terms in addition to the common linear viscous expression. In some cases a nonlinear substitute for assumed linear damping may be more appropriate. Even for exceptional cases where all nonlinearity may be ignored, it is shown that viscous dissipation involves subtleties that can lead to huge errors when ignored.

## INTRODUCTION
(Note: the use of the adjective, 'linear', in the title of this article -- is to be viewed only in the following context -- for all cases considered, the amplitude of pendulum motion is small enough for reasonable validity of the approximation $\sin\theta = \theta$.)

### Preface
Because of the complexities of friction, most models of oscillator damping concentrate on a single process. For harmonic motion at small amplitudes within a fluid, the simplest energy loss is viscous loss that is proportional to the velocity. For an oscillating sphere, it is tempting to treat this friction as simply the steady flow expression involving the viscosity of the fluid and the radius of the sphere (Stokes law). We will see that such an approximation results in huge errors. Not only must one account for 'history effects' because of the repetitive motion, but estimates of the damping generally require a full nonlinear fit to decay records. The 'least' additional friction that must be included is that which depends on the square of the velocity.

The 20th century began to reveal the importance of damping friction that is internal to the structure of an oscillator. Just as the pendulum was an important instrument for the study of fluid (external) friction, it is also vital to the study of internal friction. In studies of the pendulum, it has been common to restrict one's attention to damping that derives only from the fluid. Or if the pendulum were envisaged as swinging in a vacuum -- to ignore any processes other than those localized at the axis; which is typically either a knife-edge or a flex-pivot. Both of these approaches are naive; it will be seen in the experiments that follow, that a pendulum is generally influenced by mechanisms of both solid and fluid type, neither of which can be properly viewed as completely linear.

### Background
Before the developments of deterministic chaos, there seemed to be an obsession with the harmonic (linear) oscillator. Our understanding of more realistic macroscopic vibrations improved dramatically with enlightenment concerning the importance of elastic nonlinearity. While the developments of this period focused on elastic nonlinearity, there was little concern for a different type of nonlinearity that begs for more study--damping that is other than the classic examples which have been known either for decades or centuries.

Early experiments demonstrated clearly that the linear approximations of common damping theory were useless, when applied to certain classic motions. These include (i) moving in a fluid at higher Reynolds number, where a term that is quadratic in the velocity must be added to the viscous friction normally employed, and (ii) oscillations that involve sliding or rolling surfaces, where Coulomb's nonlinear friction must be used; which considers only the algebraic sign of the velocity. More recently, there have been considerations for (iii) the specialized (human-adjustable) types of nonlinear 'damping' that derive from feedback, as in the van der Pol oscillator. We will later in this paper demonstrate how both of the nonlinear forms of fluid and Coulomb damping are frequently present in a single free-decay of a pendulum (along with exponential damping). Nonlinear fluid damping is most important early in the decay, and Coulomb damping is most important late in the decay. On the other hand, exponential damping is most important in the middle of the decay.

By swinging a pendulum in water, we will also show that the common linear viscous theory is subject to serious misinterpretation--by assuming that the damping coefficient is a constant independent of fluid density and the frequency of oscillation. Highlighting of this issue is in large measure possible because of another element of the study, which is described in the paragraphs which follow.

The present work is largely concerned with a very important, but mostly unknown, ubiquitous form of nonlinear damping. Following the example of those who were the first to note and to label it as universal [1], the present paper will formally refer to this damping as the *'universal internal friction of solids'*. Without the adjective 'universal', this label is analogous for solids to that which was employed for fluids by the person who was first to properly describe it [2]. What we call viscous or fluid friction, Sir Stokes called 'internal friction of fluids'; moreover, what we call 'viscosity', he referred to as the 'index of friction'. Why his preferences were lost to history is a curiosity unknown to this author.

For those engineers who discovered long ago that nature refuses to let them ignore it, the internal friction of solids has been called 'hysteretic" [3]. It has even found a prominent place in a part of 'big science', where it is called 'structural damping' [4]. The last reference is to those whose passion is to directly observe gravitational waves with an interferometer. At LIGO, VIRGO, GEO600, and TAMA--mirrors hang from pendulum supports--and if it were possible, hysteretic damping would be eliminated from them; so that thermal noise of their detector would be significantly reduced.

Studies by the present author have also uncovered (at relatively low-energies of a pendulum) features of hysteretic-related damping that manifest themselves through discontinuous jumps [5]. These are reminiscent of the Portevin LeChatelier effect, which is likely related to the better known (at least to old physicists) Barkhausen effect. The damping in this regime is more properly labeled 'complex' than simply 'nonlinear'. For systems that dwell in the 'low and slow' (small energy, long period) world of a pendulum, mesoanelastic complexity can be the basis for bizarre behavior [6].

Debate continues as to whether hysteretic damping is really linear or nonlinear (since much additional study is needed for a definitive proof). Engineers have usually chosen to include hysteretic damping in their models by way of an artificial 'equivalent viscous' form. Nevertheless, because the damping can be understood in terms of secondary creep of the pendulum structure [7]; hysteretic damping is most naturally cast in nonlinear form [8]. The same nonlinear mathematical form that is well-suited to its description, will be seen to naturally accommodate, in a single equation, the other commonly encountered types of oscillator damping.

A factor in the linear/nonlinear controversy involves the shape of the decay curve--being exponential for hysteretic damping, the same as viscous damping. It should be noted, however, and as will be discussed in detail later; there is a significant difference in the frequency dependence of damping that is external (viscous), as opposed to internal to the oscillator (hysteretic). Moreover, it will be seen that the frequency dependence of linear fluid damping is significantly different from common assumptions. For a pendulum swinging in water, the Q variation with frequency was found to depart radically from what is usually assumed. Perhaps a similar result is true for motion in air, but this would be more difficult to demonstrate.

In the experimental results which follow, a heuristic 'universal' model of nonlinear damping will be presented. Output generated by a single one of its pieces shows the characteristics of exponential 'hysteretic' damping . Likewise, other individual pieces generate each of the other common forms of damping: (i) exponential linear, (ii) amplitude-dependent (nonlinear fluid), and (iii) Coulomb. For this reason, and especially because it readily accommodates the important universal form of damping in solids (hysteretic), the model is itself described as universal. Because of its significant similarity to Coulomb damping, it is also frequently referred to as a 'modified Coulomb' model.

**Compound Pendulum Experiment**

A compound pendulum was studied in free-decay. It is a pendulum supported by small, low-friction, precision ball-bearings that define the axis of rotation; and which were chosen to insure a detectable, though minimal amount of Coulomb friction. It was essential to this study that the frequency of oscillation be variable; this was accomplished by adding to the usual lower mass, a position-adjustable mass located above the axis.

At all times during the decay, position of the pendulum was determined with a highly linear and sensitive detector, which provides a measurement that is virtually noninvasive, over a large dynamic range. Use of the caveat, 'virtual', here draws attention to the fact that a perfect measurement is impossible. As discussed later in this article (damping redshift), even 'classical' measurements (in addition to well known quantum cases) are not always immune to difficulty. There is at least one example from classical physics where an attempted meaningful measurement is met instead with a signifcant (state altering) disturbance to the system which is the object of the study.  It occurs when one tries to accurately measure electrical potentials on a set of small capacitors.  Anyone who has tried to do so will quickly acquire admiration for Michael Faraday.

In the present studies, although amplitude of motion was small, relative to restoration nonlinearity; it was nevertheless large enough for nonlinear viscous damping to be visually evident in the early portions of many records.  A typical free-decay record would thus show apparent evidence for (i)  non-viscous (high Reynolds number) fluid friction in the early-part, (ii) exponential decline in the mid-part, and (iii) Coulomb decay in the end-part of the decay.

The importance of a given type of damping will be shown to depend on both (i) frequency of oscillation, and (ii) time in the free-decay at a given frequency.  At higher frequencies, where viscous friction is more important, it was discovered that the textbook form of viscous damping can lead to meaningless conclusions.  Huge errors result when one assumes the damping coefficient is a constant.  Rather (as known to some engineers), it was found that the damping coefficient depends significantly on the density of the fluid as well as its viscosity (and also the frequency of oscillation).  The dependence on density and frequency of oscillation, as well as viscosity, is a consequence of harmonic disturbance to the fluid by the oscillating pendulum.  The simple Stokes law form of viscous drag (to which the proper friction term reduces in the limit of zero frequency) applies only to steady flow.

The study also shows that hysteretic damping is the dominant source of dissipation at low frequencies of oscillation, because of structural anelasticity.  Whereas this was already known by the author to be true for the case of a pendulum swinging in air, the same result was found for motion in water.  It is doubtful that most would anticipate this surprising result.

**Data Processing**

Output from the sensor of the pendulum was converted from analog to digital form and stored in the computer for analysis.  In every trial, decay of the motion was viewed until such time as noise was of comparable size to the signal.  It was found that working with signals of abbreviated duration can result in serious misinterpretation.  Such errors of judgment are no doubt a factor in many false conclusions that appear  to characterize the world of damping science.

Output from the sensor was used to compare turning-point (pendulum amplitude) data against corresponding model generated numbers produced with a computer.  In the process of refining the method, the following conclusion was reached.  Just as a full decay record is essential, as noted above; so also, a proper interpretation of the decay requires a full-nonlinear fit to the data.  The importance of a full-fit is illustrated in the following confession by the author.  Yielding to a temptation that is probably commonplace, he tried to estimate the viscous component of damping by restricting attention to the 'tail' of the decay record.  To focus on a  portion of the record seemed reasonable, since the early and obvious influence of quadratic-in-the-velocity fluid friction has by then apparently 'died away'.  It was not realized until later, following proper treatment -- that 'fits to the tail' are highly sensitive to one's choice of  'initial' condition.  Because the starting value of such a tail-fit is ill-defined, the results led to unrealistic conclusions, which at the time seemed quite reasonable.

**Prerequisites**

This study could only be realized because of three high-technology experimental components :  (i) an inexpensive, yet powerful personal computer, (ii) an adequate, yet still user-friendly analog-to-digital (A/D) converter with powerful software support, and (iii) a recently developed, superior sensor technology.

The large number of sample points  necessary for proper evaluation of a decay record makes the computer indispensible.  It is easy to see why Stokes' appeal (in 1850) for more 'decrement of the arc' measurements in support of his theoretical analyses --probably lacked for enthusiastic volunteers.  Practically speaking, the absence of any one of the elements mentioned above would have caused the present study to be impossible.

# FLUID  FRICTION  THEORY

In his seminal 1850 paper [2], Sir George Gabriel Stokes introduced his now famous law of viscous damping. Among other firsts, the theory of Stokes showed why raindrops in clouds are able to persist--since the force of the air acting on a drop depends on the radius of the small water sphere, rather than on its surface area.  His insight corrected an error in the 'common theory' which had prevailed to that point in time. It is noteworthy that this foundational work centered on one of the oldest instruments of experimental science, the pendulum.

Discussed in nearly every textbook of physics, the simple form of Stokes law of friction force f is written

$$f = 6\pi \eta a v \quad , \tag{1}$$

which involves the radius a of the sphere, moving at constant speed v, in an incompressible fluid of viscosity η . Curiously, Stokes coined an alternative expression-- 'index of friction'--why it is obscure in the history of physics is surprising and unknown to the author.

Few recognize that the steady flow expression (Eq. 1) is inadequate when the sphere executes harmonic motion. For the harmonic cases of present interest, a 'history term' must be added to Eq. 1 to yield

$$f_{harmonic} = 6\pi \eta a (1 + C_H \frac{a}{\delta}) v, \quad \delta = \sqrt{\frac{2\eta}{\omega \rho}}, \quad (C_H \to 1 \text{ as } v \to 0) \tag{2}$$

Thus, in addition to the viscosity, the harmonic viscous friction force involves the 'penetration depth' δ; which itself depends on the angular frequency ω of oscillation, and the density ρ of the fluid [9]. Through comparisons of theory and experiment, a validation of Eq. 2 is later provided. Probably as a surprise to most, it will be shown that commonly encountered cases yield a δ that is smaller than a by roughly an order of magnitude. In other words, *viscous harmonic friction can be much greater than steady flow viscous friction !* When this is the case, the retarding force is proportional to the area of the sphere rather than being proportional to its radius.

## EQUATIONS OF MOTION

**Traditional Damped Harmonic Oscillator**

The simplest equation to approximate the free-decay of most oscillators is the following famous expression:

$$\ddot{x} + \frac{\omega}{Q}\dot{x} + \omega^2 x = 0 \quad , \tag{3}$$

where the damping term has been written in 'canonical' form by means of the quality factor Q. For systems with fairly small damping, Q is defined as 2πE/|ΔE|, where E is the energy of oscillation and ΔE is the energy lost per cycle because of dissipation. More commonly, the dissipation is expressed in terms of a damping coefficient β, where

$$2\beta \dot{x} = \frac{\omega}{Q}\dot{x} \quad . \tag{4}$$

Although the damping coefficient is oftentimes called a damping 'constant', it will be later shown, using Eq. 2, that this coefficient involves the density of the fluid and the frequency of oscillation.

Using the period T = 2π /ω, the logarithmic decrement of the motion is given by

$$\delta = \beta T = \frac{\pi}{Q} \quad . \tag{5}$$

**Highly touted, but irrelevant Damping redshift**

In textbooks, a subscript 0 is usually associated with ω in Eq. 3, because of the damping 'redshift' inherent to its mathematical solution.  For the present (and possibly every) situation, the subscript is deemed unnecesary for the reasons that follow.

To his knowledge, the author alone has seriously treated the difficulty of measuring the damping shift,  which is conveniently expressed in terms of the Q as

$$\frac{\Delta \omega}{\omega} = -\frac{1}{8Q^2} \quad . \tag{6}$$

The effect does not even exist to begin with, in the case of  hysteretic damping (discussed later), and it is too small to permit experimental validation--except, perhaps, through the employment of Herculean efforts. Consider, for example, the smallest Q encountered in the present experiments, having the value of about 10.  According to Eq. 6, the magnitude of the expected shift, in this largest of all the cases studied, is only 0.125 %.  Such a small change in frequency for a short-duration signal is difficult to measure by either of two methods that were attempted by the author with students, using the experiment that is next described.

In lieu of a pendulum, where control of the damping parameter is difficult, an RLC oscillator in free-decay was studied, using  fixed L and C, and a variable resistance R.  Individual decay records were captured with a Tektronix digital oscilloscope and then exported to Microsoft Excel for detailed analysis. The experiment yielded the following conclusions.

(i) Use of  spectral measurement techniques (involving the fast Fourier transform, or FFT) do not work for reason of the Heisenberg uncertainty principle.  As resistance is increased to enlarge the effect, 'lifetime' shortening of the waveform becomes a serious detriment; i.e., concomitant linewidth broadening disallows the shift to be resolved.

(ii) Use of  computer fits to the data prove no better, for reason of noise associated with bandwidth features of the oscilloscope.  Reducing the lifetime of the signal requires an increase in the bandwidth of the recording instrument, lest the signal become distorted.  The concomitant increase in Johnson noise becomes detrimental to this measurement.  This limitation became evident in the noise-reduced quality of residuals-- difference between data and a theory-generated best-fit to the data.

## The Pendulum--where Linear Approximations are insufficient

Eq. 3 is attractive because of its simplicity; it is especially simple as here presented, where attention to the inconsequential damping redshift has been omitted.  Because it is a linear equation, it cannot be used to describe a host of increasingly important phenomena, some of which can be studied meaningfully with the pendulum [10].  The present work demonstrates that nonlinearity is important not only to pendulum restoration, but also to pendulum damping.   Whereas pendulum damping of nonlinear viscous type is fairly well known,  other important nonlinear types are not.

**Large -amplitude  Pendulum**

The driven, rigid simple pendulum with linear damping has become an archetype of chaos [11]. The equation of motion for its angular displacement θ, when driven by an external torque $\tau_d$ at frequency $\omega_d$ is

$$I\ddot{\vartheta} + \frac{\omega I}{Q}\dot{\vartheta} + Mg\Delta \sin\vartheta = \tau_d \sin\omega_d t \quad , \tag{7}$$

where I is the moment of inertia, and Δ is the distance of the center of mass from the axis.  The restoration is seen to be nonlinear, because of the sine term.  It is necessary for chaos.  The damping term is linear, being proportional to the angular velocity.  For the case of Mercer's online chaotic pendulum, this is a reasonable approximation, because most of its damping derives from eddy currents of an aluminum disc that is connected to the pendulum, and which moves between magnets.

**Small-Amplitude Pendulum in Free-decay**

Unlike pendulum motion-past-vertical, which is necessary for chaos; all present experiments involved amplitudes sufficiently small that sin θ may be replaced by θ in Eq. 7. (We will see that the elastic nonlinearity remains observable (using residuals) for motion at some of these lower levels; but its magnitude is not consequential, as compared to damping nonlinearity.) Additionally, the pendulum was studied in free-decay, so that $\tau_d$ of Eq. 7 is zero, and the equation in turn reduces to the equivalent of Eq. 3.

### Heuristic Universal form of Pendulum (Mechanical Oscillator) Damping

What has been discovered through present and previous experiments by the author, is that Eq. 3 is rarely an adequate description of pendulum motion, even at amplitudes where concerns for isochronism is not an issue. To correct the observed deficiencies, the following nonlinear equation was developed.

Through a large number of tests, involving various types of pendula with differing dissipation types, the efficacy of the following equation has been demonstrated (c.f., reference 8). Later in this document, the usefulness of Eq. 8 for deciphering experimental observations will be further illustrated with some examples.

$$\ddot{x} + [\frac{\pi\omega^2}{4Q_{c0}} + \frac{\pi\omega}{4Q_h}\sqrt{\omega^2 x + \dot{x}^2} + \frac{\pi}{4y_0 Q_{f0}}(\omega^2 x + \dot{x}^2)]\text{sgn}(\dot{x}) + \frac{\omega}{Q_v}\dot{x} + \omega^2 x = 0 \quad . \quad (8)$$

Of the four damping terms in Eq. 8, the last one is traditional (linear) viscous, and the three inside the brackets are nonlinear because of the signum (sgn) function which operates on the velocity. Respectively, from left to right, the bracketed terms model friction of the type that is: (i) Coulomb, (ii) hysteretic, and (iii) amplitude dependent (or 'fluidic'). As presented, both the hysteretic damping and the amplitude dependent damping can be justifiably labeled with the generic term, 'modified Coulomb' damping. They may be thought of as a friction force where the 'coefficient' is energy dependent, as opposed to being the kinetic constant used to treat contacting surfaces where there is relative, sliding motion.

It is important to note that only hysteretic damping and viscous damping are characterized by a constant Q. Also, in the case of amplitude-dependent damping of Eq. 8-- although it may be called 'fluidic', because of similarity to nonlinear fluid damping; the terminology can be misleading. Fluid viscosity has no physical basis for those cases where the dissipation involves the rearrangement of crystalline defects in the structure of the solid.

It is also worthy of note that each of the bracketed parts can be written as a force per unit mass, in terms of the energy as

$$\frac{f}{m} = c\left(\frac{2E}{m\omega^2}\right)^\lambda = \frac{\pi}{4}\frac{\omega}{Q}[v] \quad . \quad (9)$$

where c is a constant. For Coulomb damping λ = 0, for hysteretic damping λ = 1/2, and for fluidic damping λ = 1. Some yet-to-be discovered damping type may take advantage of an altogether different value of λ.

It is also seen from Eq. 9 that the friction can be expressed in terms of a square wave velocity, [v], which corresponds to the fundamental component of a Fourier series expansion of the oscillator's actual velocity. (For a square wave ±h, the amplitude of the fundamental is ±(4/π)h.) We thus see that all damping which involves harmonic oscillation, when expressed in canonical form, is of the form f = mωv/Q. The simplicity of this result is probably why viscous damping has been viewed by so many as 'inviolate'. One must exercise care, however, because only for the case of hysteretic and viscous damping is Q constant. If we designate the amplitude of the motion by y, and indicate initial values with the subscript 0; then for amplitude-dependent damping $Q_f = Q_{f0}(y_0/y)$. For Coulomb damping $Q_c = Q_{c0}(y/y_0)$. Thus the Q of hysteretic damping increases with time, whereas the Q of Coulomb damping decreases wtih time.

The net Q of the oscillator, when all damping types are simultaneously active, is expressible in the form

$$\frac{1}{Q} = \frac{1}{Q_c(t)} + \frac{1}{Q_h} + \frac{1}{Q_f(t)} + \frac{1}{Q_v} \quad . \tag{9}$$

This equivalence to the manner in which capacitors in series combine should not be surprising; the smallest Q of the system is most important to the energy loss.

The primary difference between hysteretic and viscous damping involves their frequency dependence. It is not difficult to show from Eq. 8, that for hysteretic damping

$$Q_h \propto \omega^2 \quad \text{(hysteretic)} \tag{10}$$

It is popularly and erroneously believed that the Q of viscous damping is always proportional to the first power of the frequency. This would be true if the damping coefficient were constant, which it is not. In the water-damping results which follow, it was found that the exponent was different according to which configuration of the pendulum was used

$$Q_v \propto \omega^x, \quad x \approx -5.6, \text{ high freq.}$$
$$, \quad x \approx 0.50, \text{ low freq.} \quad \text{(linear viscous)} \quad , \tag{11}$$

The low frequency configuration used a pair of weights on the pendulum rod--one above, as well as below the axis of the pendulum; whereas the low-frequency configuration used only the lower weight.

Eq. 11 illustrates a confusion factor associated with the common method for specifying internal friction; i.e., as 1/Q. The quality factor of an oscillator formally has meaning only as it is presently defined. The Q involves more than just the 'loss tangent' associated with stress-strain hysteresis; its frequency variation also involves restoration properties of the oscillator. Although many experimenters choose to designate their measured loss tangent by 1/Q, the practice is misleading. The resulting confusion for hysteretic damping of a pendulum is discussed in detail in [12].

The functional form of Eq. 10 is especially significant to two modern experiments: (i) torsion pendulum measurements of Newton's universal gravitational constant, G [c.f., 13], and (ii) LIGO's search for gravitational waves, in which a given mirror of the interferometer is supported as a high Q pendulum. In [4] one reads the following statement: "The quality factor (Q) of the IP is compatible with structural damping. Q is proportional to the square of the pendulum frequency".

## Basis for Curve Fits

For purpose of computer generated curves to compare theory with experiment, it is necessary to obtain a mathematical representation, consistent with Eq. 8, for the turning points of the pendulum. Such an expression is generated from the time dependence of pendulum energy [8]:

$$\dot{E} = -(c_1 + c_2\sqrt{E} + c_3 E)\sqrt{E} \quad , \tag{12}$$

where each of the C's is a constant. Both hysteretic and viscous damping are accommodated by the constant $C_2$. Any non-zero Coulomb damping is represented by $C_1$, and $C_3$ is for any amplitude-dependent damping that might be present ( high Reynolds number cases, or some types of internal friction in solids ). Superposition of the energy parts is used in writing Eq. 12, even though nonlinear mathematics does not in general permit it.

Eq. 12 provides the first order differential equation for the turning points (corresponding to maximum displacement = amplitude y), by recognizing that energy is proportional to $y^2$. Thus, Eq. 12 yields

$$-\dot{y} = ay^2 + by + c \quad , \tag{13}$$

where the three constants a, b, and c are various mixtures of the constants of Eq. 12.. The choice of symbols a, b, and c allows convenient association with the well-known positive root of the quadratic equation of algebra. Thus the solution to Eq. 13 is obtained from integral tables in terms of $\mathrm{r} = \sqrt{b^2 - 4ac}$ as

$$y = \frac{b(p-1) + \mathrm{r}(p+1)}{2a(1-p)} \quad , \quad p = \frac{2ay_0 + b - \mathrm{r}}{2ay_0 + b + \mathrm{r}} e^{-\mathrm{r}t} \quad , \tag{14}$$

where it is required that r be real (thus $c \leq b^2/4a$). Although an alternative solution could be written for cases of large Coulomb damping (determined by c), Eq. 14 is the appropriate solution for present purposes. Amplitude-dependent damping is contained in the constant a, and hysteretic and/or viscous (linear) damping are contained in b.

## Quality Factor

Once the parameters a, b, and c of the decay have been estimated, using the method described later, the Q of the oscillator (time dependent for non-zero a and c) is calculated from the frequency f using

$$Q = \frac{\pi f}{(ay + b + \frac{c}{y})} \quad . \tag{15}$$

## EXPERIMENTAL APPARATUS

The pendulum used in these studies is pictured in Fig. 1. As noted earlier, bearings were purposely selected, instead of a knife edge for the pendulum axis, to provide minimal Coulomb friction, in addition to other types of damping normally active in the pendulum. The pair of ball-bearings (System Management Associates, Inc, Box 957, Palo Alto 94302, part No. 628) are located in the arms of a y-shaped plastic housing that is mounted with a bolt to the gear-moveable platform (via the knurled knob), which rides on the support post of square cross-section. (For present experiments, this geared degree of freedom was not needed).

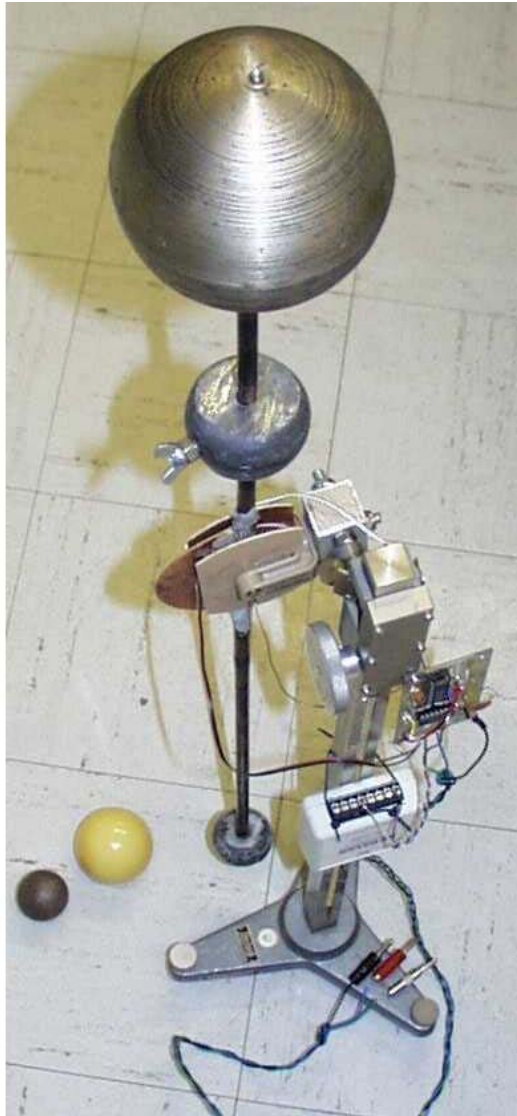

**Figure 1.** Compound pendulum showing five different masses that could be used in a variety of configurations.

In addition to the dark-colored ping-pong ball and the cue-ball resting on the floor, three other masses are present in the picture, mounted on the rod. At the bottom of the rod is a nearly spherical lead mass which can be attached and detached with a screw. At the top is a detachable light-weight, large-diameter hollow metal ball; and between this pair is a lead weight (711 g) whose position can be changed by means of its wing-nut clamp. Not all the objects shown in the figure were used in the data presently described.

      The hollow rod which holds the masses was once a lightweight aluminum alloy arrow, whose total mass with additional components is now 44.4 g, and whose length is 77 cm. The fletching and end-nocking plastic pieces were cut away from the arrow, and it was squeezed flat using a vise, in a central region about 10 cm long, except for a small section at the geometric center. At this center, a through-hole was drilled to accommodate the small brass pin, whose ends are located (operationally) in the inner race of the ball bearings. After assembly, this pin was glued in place to the rod.

      A semicircular piece of single-sided printed circuit board was also epoxied to the flatted portion of the rod to serve as the moving electrode of the sensor. To strengthen the hollow rod at the transition between its circular and flat regions, epoxy was added. To insure an equipotential throughout the moving electrode that comprises PCB and flatted section of the arrow, paint was scraped from the arrow and silver paint was added to insure electrical contact between the arrow and the copper of the board.

**Sensor**

The photograph of the pendulum in Fig. 2 is a closeup which focuses on the symmetric differential capacitive (SDC) sensor and its electronics support [14].

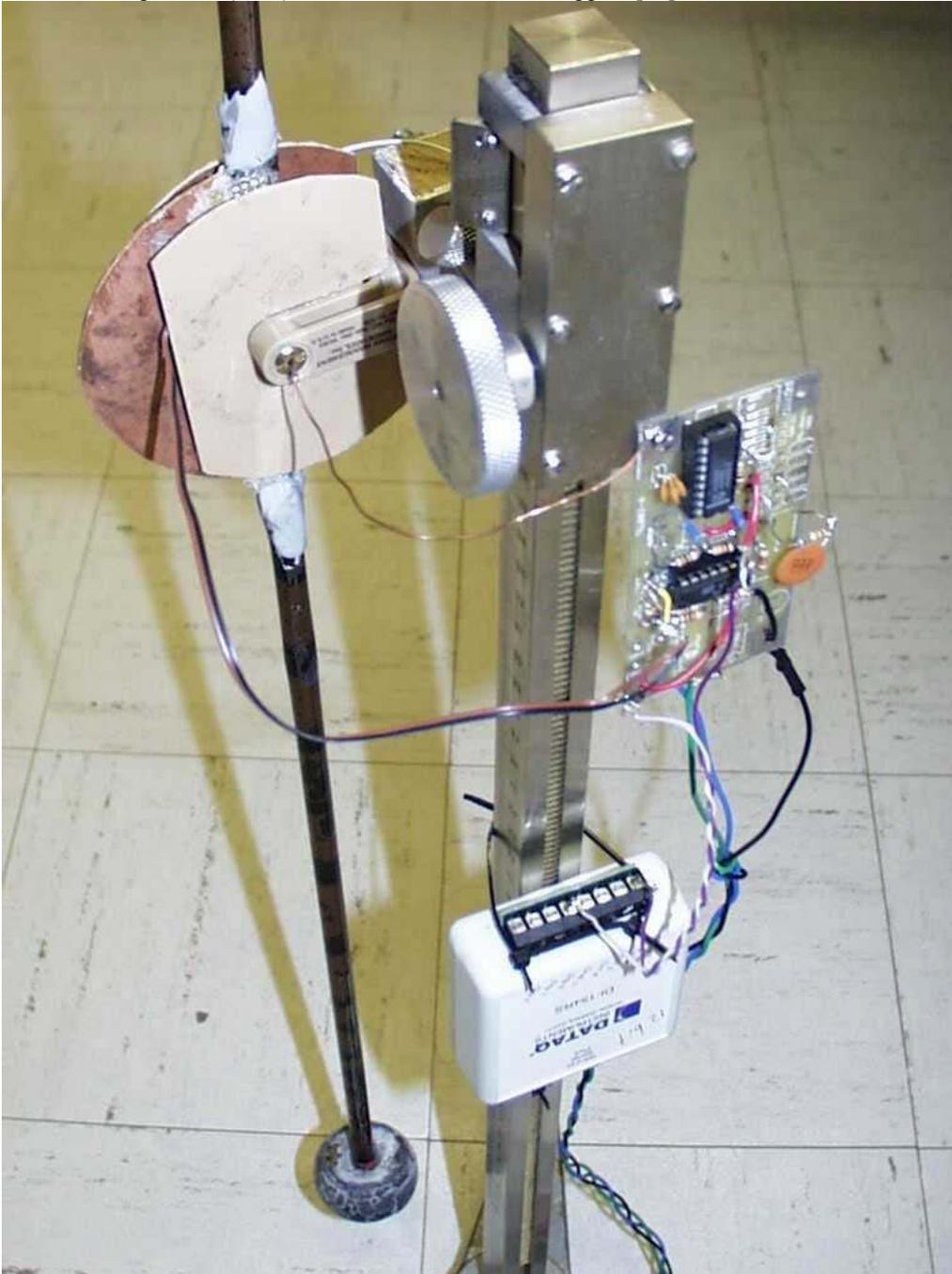

**Figure 2.** Closeup of the sensor and its support electronics.

Clearly visible in Fig. 2 is one of the pair of static electrodes of the sensor, which are positioned on opposite sides of the rotating electrode. These are not the full circle type that would permit angle measurement of $\pm \pi/2$; since the plastic part which holds the pair of ball-bearings disallows motion larger than about $\pm 0.7$ rad.

On the side of the post opposite the sheet copper electrodes, and supported by a single small screw at its upper left, is the circuit board (dim. 4.5 cm x 9 cm) on which are installed electronic components used by the sensor. These include a quad-opamp and NE5521N integrated circuit chips. The banana plugs through which electrical power is supplied to the board from a power supply (not shown) are visible in Fig. 1.

Seen also with the sensor in Fig. 2 is a Dataq DI154-RS analog-to-digital converter. Being limited to 12-bit resolution, this converter was replaced for present experiments with a 16-bit unit (not shown, Dataq's DI-700) that operates through the USB of the computer.

Also not shown in the figures is the tall nylon shroud that was used to shield the instrument from air currents when operating at long periods, where the sensitivity is great. This shroud was fabricated from two barrels, one of which had both ends removed using a skil-saw. The 2nd, half-length barrel sits on top of the open-end cylinder; which itself surrounds the pendulum, which during operation, sat on a sturdy laboratory table.

**Sensor Calibration Results**

Sensor linearity is a requirement of great importance for work of present type. Shown in Fig. 3 are the results of calibration (small displacement data), which clearly demonstrate that the requirement was met for the range considered by the figure. Although the sensor was crudely constructed, its linearity was found to be excellent over the full range of displacements considered in the study. The precision data of Fig. 3 were obtained with a digital vernier caliper. Although it was not possible to calibrate the sensor with this great a precision at angles beyond those of the figure, the indicated 3.85 V/rad is stated with confidence for all data taken.

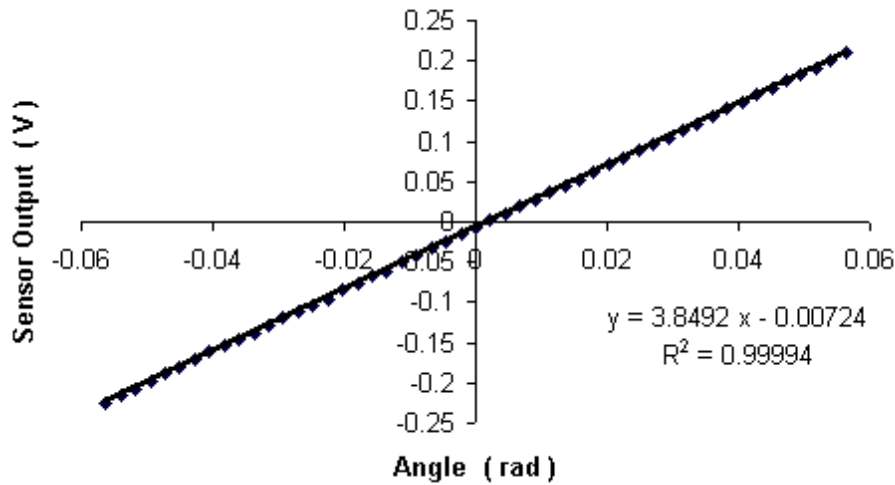

**Figure 3.** Sample calibration results for the capacitive sensor. The detector's response was determined to be constant at 3.85 V/rad over $\pm$ 0.6 rad (larger than the range of the figure by an order of magnitude).

## EXPERIMENTAL RESULTS

### Benchmark of Internal Friction -- Low Frequency Motion in Air

Even Stokes recognized that internal friction of the pendulum could not be ignored, and he talks about the matter in his 1850 paper [2]. For the simple pendulum, internal friction will not be as significant as for a compound pendulum like the present instrument. An attempt to 'benchmark' the hysteretic component is important for the identification of damping types (similar, in principle, to determining background in nuclear experiments).

Numerous investigations by the author and by others have shown that internal friction of hysteretic type is the dominant source of dissipation in any long-period mechanical oscillator. A seminal paper might have prevented this important fact from having to be rediscovered--had the 1927 work of Kimball and Lovell become better known [1]. The frequency dependence of pendulum Q for hysteretic damping (noted in Eq. 10 ) is consistent with the 1927 discovery.

Kimball and Lovell used neither a pendulum nor any other type of oscillator for their work; rather they devised a clever technique which measured the transverse deflection of a rod. When hysteresis due to anelasticity is present, a rotating rod will experience, in addition to the primary deflection which results from an end-applied bending force--an additional deflection that is transverse to the first. In showing that the resulting transverse deflection was independent of the rate of rotation (and thus frequency), they were first to determine that the internal friction of solids is functionally different from the internal friction of fluids. Stokes was the first to properly describe the internal friction of fluids 77 years earlier. Kimball and Lovell chose to label the damping they had discovered as 'universal' to solids, and their claim has been substantiated by recent experiments. Perhaps this choice caused them to be perceived by their peers as arrogant, so that the paper was poorly received. Whatever the cause, their work did not receive the prominence that it deserves, until the 1990's.

Shown in Fig. 4 are the present benchmark results, which are seen to be in agreement with hysteretic (internal friction) damping theory, as evidenced by the straight line of Q versus the square of the frequency. As noted earlier, one is not forced to assume a nonlinear damping model, to obtain agreement of the type provided by the figure. An equivalent viscous damping model would also suffice, where one artificially divides the usual viscous damping coefficient by $\omega$. To do so is questionable, however; since its physical basis is unjustified. By contrast, the proper frequency dependence is automatic with Eq. 8.

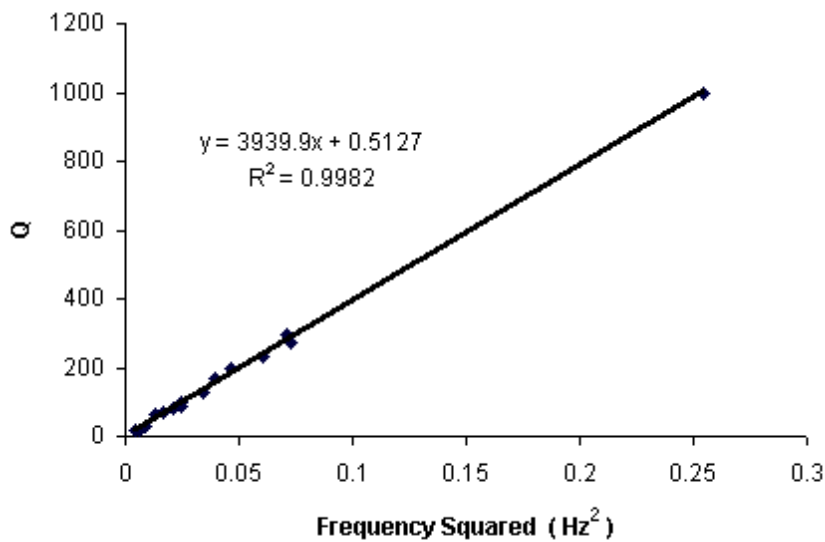

**Figure 4.** Illustration of hysteretic damping of the pendulum, for oscillation at low-frequency in air.

For the data of Fig. 4, the pendulum was swung with two masses--the pair shown mounted on the pendulum in Fig. 1, but with the large hollow sphere removed. The highest frequency included in the figure corresponds to a period of approximately 2 s. This is a shorter period than would normally still correspond so strongly to internal friction damping--except that here the pendulum rod is relatively weak for the size of weights employed (and thus deflects significantly). If the rod were more rigid, viscous damping would already have shown its presence, once the period had been adjusted downward from 15 s toward 2 s.

Even a simple rod pendulum, constructed of malleable metal, can also be configured to exhibit damping of the Fig. 4 type, as shown in reference 10. Deviations from the smooth decay curves of the present study are also possible, when working with a ductile metal. Similar bizarre behavior is encountered when one studies pendulum motion at very small amplitudes, in the realm of mesoanelastic complexity [5].

## Oscillation in Water

For these experiments, a wooden sphere (dia 3.79 cm, mass 35.8 g with screw) was attached to the bottom of the pendulum rod with a 10-32 screw that had been reduced in a lathe to a small diameter for the portion that moves in water. A rectangular plastic container (11 cm x 23 cm x 5 cm) was placed and filled with enough water for the sphere to remain completely submerged in its motion along the direction of the long axis of the container. Physical constraints disallowed the path of the motion to be centered with respect to the 11 cm dimension; rather the sphere was offset by 1.5 cm.

The bottom lead sphere pictured in Fig. 1 had been replaced for these measurements with another lead weight (993 g) that could be clamped (like the upper lead weight) at any position on the rod below the axis. This change was made so that a smooth progression in frequency, from low to high, could be accomplished by adjusting both weights.

### Estimating the influence of the Viscous Damping Coefficient $\delta$

To estimate the damping coefficient $\delta$, Equations 2 and 3 were first combined (after rewriting in the form appropriate to a pendulum). This gives the following equation for the Q of oscillation, as determined solely by linear viscous damping. (We will later discuss the manner in which the viscous part can be extracted from a full nonlinear fit to the data.)

$$Q_v = \frac{I\omega}{6\pi\eta a(1+\frac{a}{\delta})L^2} \quad , \quad \delta = \sqrt{\frac{2\eta}{\omega\rho}} \quad . \tag{16}$$

Here L = 45.2 cm (distance from axis to center of sphere of radius a = 1.90 cm), and I is the moment of inertia. Governed by the position of the weights on the rod (same being true of the center of mass); there is a unique moment for each of the fifteen frequencies that were considered in the water study. The value of I for a given $\omega$ was determined by both (i) direct computation using measured distances for the known weight masses, and (ii) indirect theoretical estimation, accounting for water influence on frequency through the effects of 'buoyancy' and 'added mass' [9]. Reasonable self-consistency was found for the two methods. Fig. 5 shows the importance to the calculation of buoyancy and added mass, whose influence on frequency is large because of the large density of water.

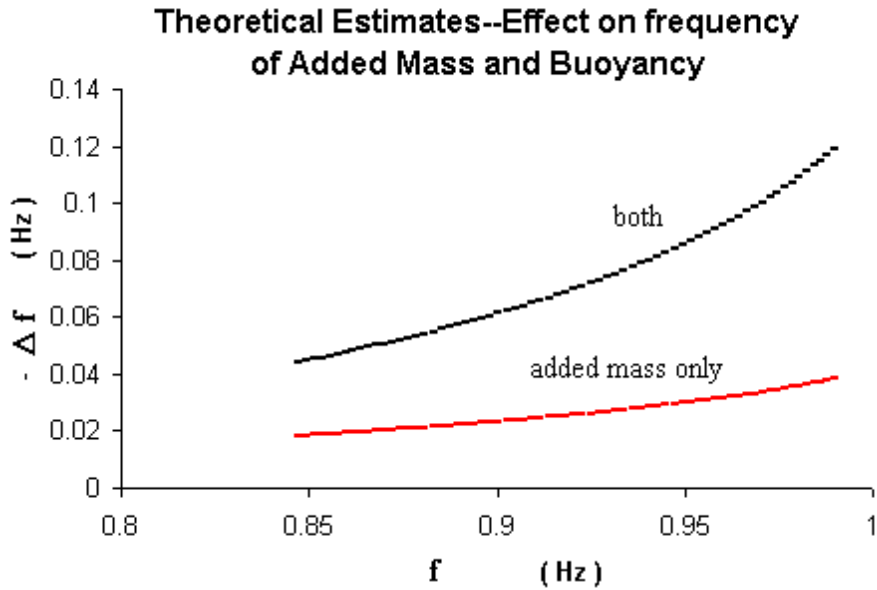

**Figure 5.** Theoretical estimate for the influence on pendulum frequency (period lengthening) due to buoyancy and added mass of the water.

The dramatic importance of the history term a/δ of Eq. 16 is illustrated in Fig. 6. To assume simple Stokes law viscous damping (which applies only at zero frequency) is to underestimate the damping by 1000 to 3000 %.

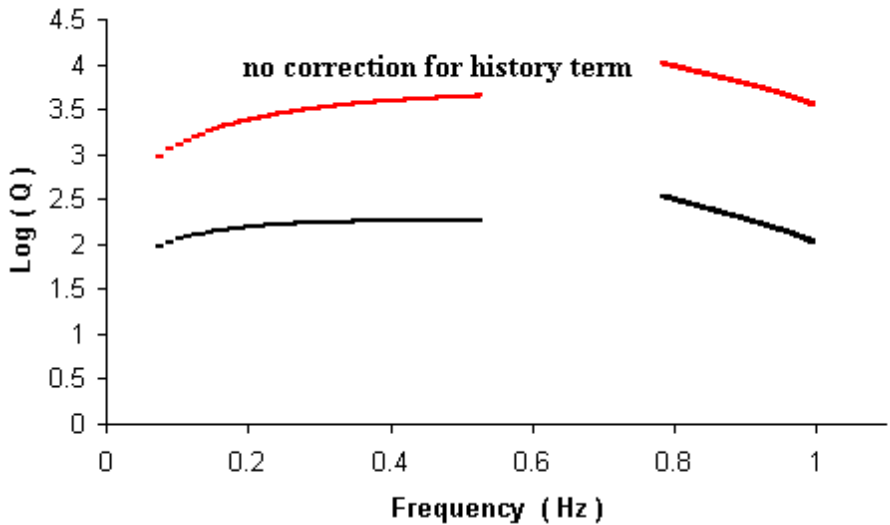

**Figure 6.** Theoretical estimates for the viscous damping limit for pendulum motion in water, with and without correction. The curves of Log Q (base 10) are for the two frequency branches considered in the study.

**Decay Records**

Of the fifteen decay records of the water study, a typical one is provided in Fig. 7; where the displayed ordinate is in terms of sensor output voltage. With the sensor calibration constant of 3.85 V/rad, it is recognized that the initial amplitude of the motion for this decay was 68 mrad.

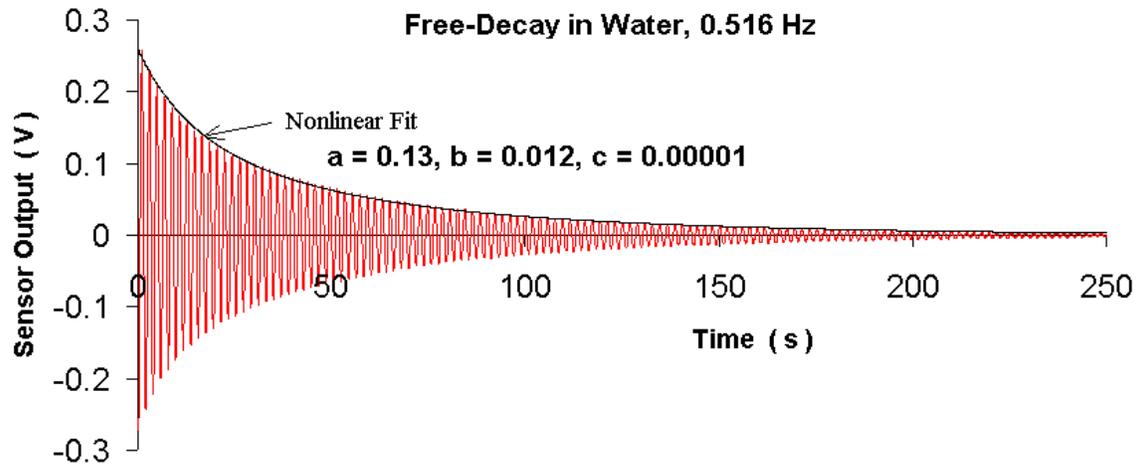

**Figure 7.** Typical free-decay record showing the influence of nonlinear damping.

The figure was generated by 'copy and paste' from Excel to Paint, after the nonlinear curve fit had been managed in Excel. The A/D converter data had been originally stored in Dataq's software package, windaq32.exe. From there it was saved in spreadsheet (*.csv) format for export to the Excel spreadsheet.

Spreadsheet generation of the fit was accomplished by 'eyeball' optimization, through trial and error adjustment of the three parameters a, b and c. This proved user friendly and powerful for Pentium II and later computers, only because of the 'autofill' function. By changing a number at the last row (bottom) of entries, and autofilling upward to the top, the effect of any change can be immediately assessed at the instant the graph is refreshed on the monitor. The process slows down with larger data sets, but some of the present cases involved 32K voltage values; which is the maximum file size for graphing in Excel. [Caution: autofilling downward can experience a 'sensitive dependence on initial conditions'--the generation of blank rows at a rate which approaches infinity! ]

## Estimating the Viscous (linear) contribution to the Damping

In Fig. 8 we see the dissipation history of the decay record of Fig. 7.

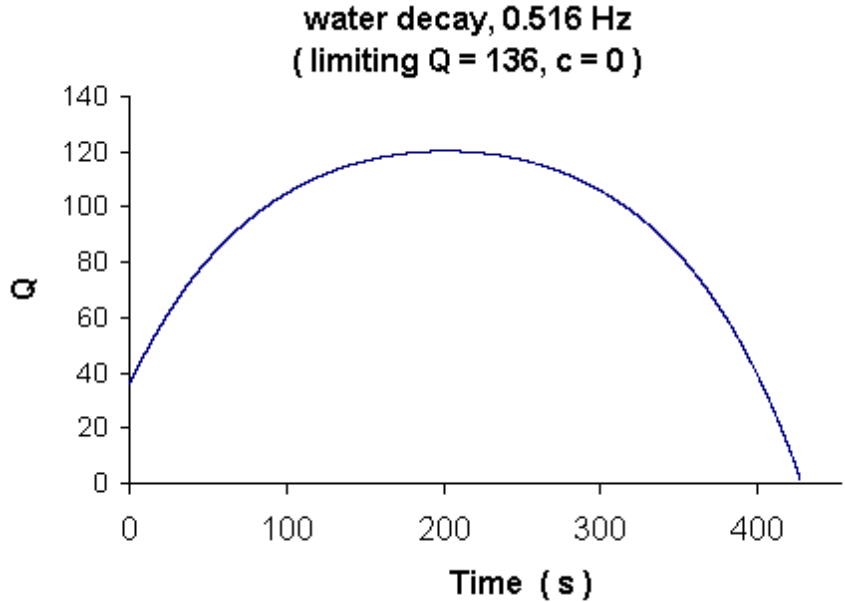

**Figure 8.** Illustration of Fig. 7's time-dependent dissipation, due to nonlinear damping.

The influence of the nonlinear damping (through parameters a and c) is readily evident in Fig. 8. In the early part of the decay, loss is greater than exponential (due to non-zero a), causing the Q to increase from below 40 toward the maximum of about 120. If there were no Coulomb damping (bearings that support the pendulum), the Q would limit at the value of 136 (obtained by setting c = 0 in the spreadsheet algorithm).

The method just described is the only dependable means known to the author for estimating the exponential (constant Q) contribution to the damping. As was noted earlier, attempts to 'fit to the tail' of the decay are fraught with subtle difficulties of non-reproducibility, even if there were not an obvious and ultimate Coulomb fall-off, as in the present case.

It should be noted that the limiting Q (136) is not necessarily viscous-only damping, since hysteretic damping also gives an exponential (constant Q) decay. To compare present results with Eq. 16, requires a correction to the raw data, due to its internal friction (hysteretic) component. The hysteretic component had been determined earlier; i.e., $Q_{hys} = 3940 \, f^2$ (refer to Fig. 4). Thus,

$$\frac{1}{Q_{v,\exp}} = \frac{1}{Q_{meas}} - \frac{1}{Q_{hys}} \quad . \tag{17}$$

Using Eq. 17, we thus adjust the measured Q upward to yield our estimate for the viscous Q at the frequency of 0.516 Hz ; i.e., 156. From Eq. 16, for the moment of inertia appropriate to 0.516 Hz ( I = .099 kg m$^2$ ), we obtain a value that is 13 % larger, at 176. This is evidently a systematic error of unknown origin (using for the viscosity of water 1.0016 mPa s and for the density 1000.0 kg/m$^3$) , since most of the 15 measurement results are smaller by roughly this amount.

Shown in Fig. 9 are the comparions with theory for all 15 cases. Here, rather than correcting the measurements upward to compare with theory, we have instead elected, in Fig. 9, to correct Eq. 16 downward to compare directly to the experimental values.

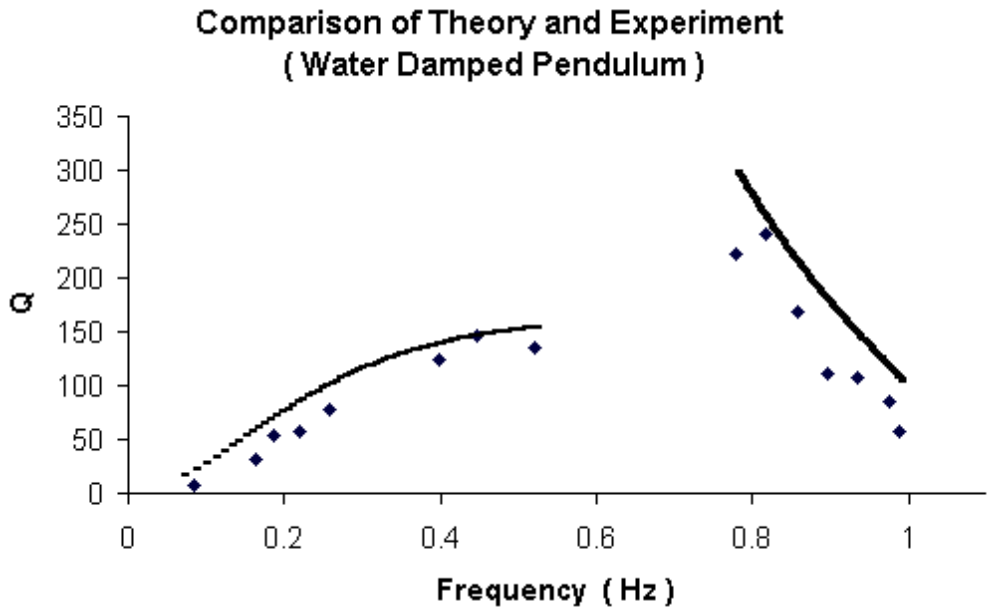

**Figure 9.** Comparison of theory against experiment. The solid curves of theory were generated by combining with Eq. 16 the Hysteretic influence detailed in the data of Fig. 4.

In generating the experimental data of Fig. 9, only the exponential (limiting) Q's were considered. Following nonlinear curve fits to each of the decay records, this limit was obtained by setting parameters and c of the fit to zero (in Eq. 15 ). The limit corresponds to a combination of low-level (linear) viscous damping and hysteretic (internal friction) damping. The latter is more important in the low-frequency branch of Fig. 9.

To ignore the hysteretic damping in the lower branch of Fig. 9 is to overestimate the Q substantially, as seen in Fig. 10.

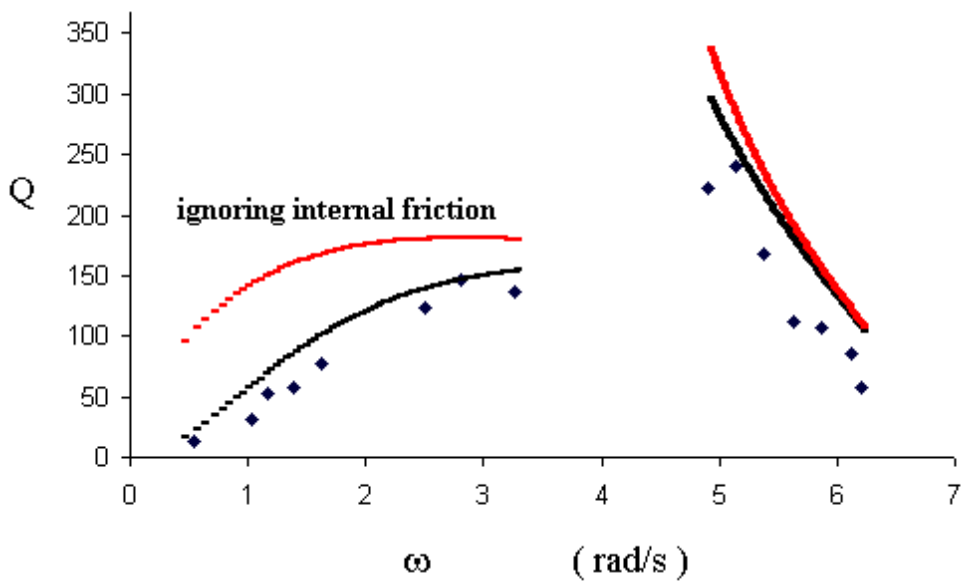

**Figure 10.** Illustration of large errors that result from ignoring hysteretic damping of the pendulum at lower frequencies (red curves).

## CONCLUSIONS

The present work, plus the independent investigation of others, provides strong evidence for a broadly important form of internal friction damping that has been ignored for decades following the earliest scientific evidence for its existence. Failure for its recognition probably derives from the fact that its most common embodiment is cause for exponential decay, just like linear viscous damping. Yet in the celebrated work of George Gabriel Stokes, written in the mid-nineteenth century, there are tell-tale indicators of its presence. Most important is the little-known 1927 paper which made claims for this universal form of damping in solids, that is not consistent with the physics of fluid friction. In spite of these strong early indicators, only since the 1990's has this universal form of damping been seriously studied.

The methods of pedagogical physics appear to have been a factor in our oversight. From their textbooks, students are led to believe that external-to-the-oscillator (viscous) friction is the most important (if not only) mechanism responsible for energy loss of an oscillator. To make matters worse, the textbook treatment of damping is not even adequate to the treatment of harmonic fluid friction. There is undue attention to what may be a non-measurable attribute of the theory provided; and students can be easily led to believe that the viscous damping coefficient is a constant, involving only the viscosity of the fluid. Although many cases of damping can be adequately described by a coefficient multiplying the velocity, it is not permissible to treat this coefficient as being independent of either (i) the frequency, or (ii) the density of the fluid in which the oscillation occurs.

Deterministic chaos took on meaning, only after our obsession with linear elasticity was abandoned; so likewise, it is felt that additional important discoveries await the elimination of our obsession with linear damping.